\documentclass[12pt]{article}
\usepackage{amssymb}
\usepackage{amsmath}
\usepackage{setspace}
\usepackage{placeins}
\usepackage{natbib}
\usepackage{caption}
\usepackage{verbatim}
\usepackage{rotating}
\usepackage{geometry}
\usepackage[utf8]{inputenc}
\usepackage{pgfplots}
\usepgfplotslibrary{groupplots}

\pgfplotsset{compat=newest} 

\begin{document}


\title{Data-driven retrieval of primary plane-wave responses}
\author{Giovanni Angelo Meles, Lele Zhang, Jan Thorbecke, Kees Wapenaar and Evert Slob}
\date{\today}
\maketitle
\begin{flushleft}   
\footnotesize{Department of Geoscience and Engineering, Delft University of Technology, Stevinweg 1, 2628 CN Delft, The
Netherlands\\}
\footnotesize{G.A.Meles@tudelft.nl}
\end{flushleft}

\section{Abstract}

Seismic images provided by reverse time migration can be contaminated by artefacts associated with the migration of multiples.
Multiples can corrupt seismic images, producing both false positives, i.e. by focusing energy at unphysical interfaces, and false negatives, 
i.e. by destructively interfering with primaries.
Multiple prediction / primary synthesis methods are usually designed to operate on point source gathers, and can therefore be computationally demanding when large problems are considered.
A computationally attractive scheme that operates on plane-wave datasets is derived by adapting a data-driven point source gathers method,
based on convolutions and cross-correlations of the reflection response with itself, to include plane-wave concepts. 
As a result, the presented algorithm allows fully data-driven synthesis of primary reflections associated with plane-wave source responses.
Once primary plane-wave responses are estimated, they are used for multiple-free imaging via plane-wave reverse time migration.
Numerical tests of increasing complexity demonstrate the potential of the proposed algorithm to produce multiple-free images from only a small number of plane-wave datasets.

\section{Introduction}

Most standard processing steps, e.g. velocity analysis (\cite{Yilmaz2001}) and reverse time migration (\cite{whitmore1983iterative,mcmechan1983px,Zhu1998, Gray2001,mulder2004comparison}),
are based on linear (Born) approximations, for which multiply scattered waves represent a source of coherent noise.
When linearized methods are employed, multiples should then be suppressed to avoid concomitant artefacts.
Free-surface multiples particularly affect seismic images resulting from marine data (\cite{Wiggins1988}), 
and many algorithms have been designed to attenuate the presence of free-surface multiples (for a comprehensive review see \cite{Dragoset2010}).
On the other hand, internal multiples strongly contaminate both land (\cite{Kelamis2006}) and marine data (\cite{VanBorselen2002}). 
Fewer techniques have been designed to estimate and remove internal multiples.
The seminal method by \cite{Jakubowicz1998}  uses combinations of three observed reflections to predict and remove internal multiples. 
However, this scheme requires prior information about reflections to allow proper multiple prediction and removal. 
On the other hand, applications of inverse scattering methods (\cite{W7})
can predict all orders of
internal multiple reflections with approximate amplitudes in one
step without model information (\cite{ten2002prediction,loer2016relating,zhang2019data}).

Multiple-related artefacts can also be dealt with via Marchenko methods.
Marchenko redatuming estimates Green’s functions between arbitrary locations inside a medium 
and real receivers located at the 
surface (\cite{broggini2012focusing,Wapenaar2012b,Wapenaar2014,DaCostaFilho2014}).
In Marchenko redatuming, Green's functions are estimated using reciprocity
theorems involving so called `focusing functions', i.e. wavefields which achieve focusing properties in
the subsurface (\cite{Slob2014}). In contrast to seismic interferometry, Marchenko redatuming requires an estimate of the direct wave from the virtual
sources to the surface receivers, only one sided illumination of the medium and no physical
receivers at the position of the virtual sources (\cite{broggini2012focusing,Wapenaar2014}). 
Focusing functions and redatumed Green's functions can provide
multiple-free images directly (\cite{Slob2014,Wapenaar2014}). 
Moreover, combining Marchenko methods and convolutional interferometry
allows estimating internal multiples in the data at the surface (\cite{Meles2015,da2017elastic}).
Other applications of the Marchenko method include microseismic source 
localization (\cite{behura2013imaging, van2017marchenko,Brackenhoff2019SE}), inversion (\cite{slob2014data,Neut2018}), homogeneous Green's functions retrieval (\cite{Reinicke2019, Wapenaar2018})
and various wavefield focusing techniques (\cite{Meles2019}).
Despite its requirements on the quality of the reflection data, and more specifically its frequency content, 
the Marchenko scheme has already been successfully applied to a number of field datasets (\cite{10.1093/gji/ggv528, van2015practical,jia2018practical,CostaFilho2017,Staring2018GEO,zhang2019field}).
Further developments have also shown how  a successful Marchenko redatuming can be achieved either via correct deconvolution of the source wavelet from 
the measured data or by including wavelet information in the Marchenko 
equations (\cite{ravasi2017rayleigh, Slob2017,Becker2018}).
Recent advances in Marchenko methods led to revised derivations which resulted in fully data driven demultiple / primary synthesis algorithms (\cite{van2016adaptive,Zhang2019,Zhang2019a}.
Different from standard Marchenko applications, in these revised derivations the focusing functions are projected to the surface, 
thus leading to the retrieval of specific properties of reflections responses in the data at the surface (i.e., internal multiples/primaries) instead of redatumed Green's functions. 
We refer to the class of applications introduced by \cite{van2016adaptive} and \cite{Zhang2019} as to `data domain Marchenko methods'.

Inspired by work on areal-source methods for primaries (\cite{rietveld1992}), Marchenko redatuming and imaging schemes were recently adapted to include plane-wave concepts (\cite{MelesVirtual}).
Here, we follow a similar approach and extend the applications of data domain Marchenko methods, originally derived for point sources, to plane-wave sources.
The benefit of using plane-wave data for imaging, i.e. an overall reduction in the data volume and the possibility to get subsurface images by migrating fewer plane-wave gathers 
than shot gathers (\cite{dai2013plane,wang2018fast,stoffa2006plane,Schultz1978}) is then combined with a fully data-driven demultiple scheme.

 \section{Method and Theory}
 \subsection{Data domain Marchenko method}
 \label{introductive_paragraph}
In this section we briefly summarize the primary reflections retrieval algorithm recently proposed by \cite{Zhang2019} and in sections \ref{horizontal_paragraph} and \ref{dipping_paragraph} 
discuss how it can be extended to include plane-wave concepts.
First, we briefly introduce the definitions and properties of the so-called Marchenko focusing functions, upon which the work on \textit{projected} focusing functions is based.
Following standard notation, we indicate time as $t$ and the position vector as $\textbf{x}=(\textbf{x}_H,z)$, where
$z$ stands for depth and $\textbf{x}_{H}$ for the horizontal coordinates $(x,y)$.
An acoustically transparent acquisition boundary $\partial \textbf{D}_0$ is defined at $z_0 = 0$ and points in $\partial \textbf{D}_0$ are 
denoted as $\textbf{x}_0 = (\textbf{x}_H,z_0)$. Similarly, points 
along an arbitrary horizontal depth level $\partial \textbf{D}_i$ are indicated as $\textbf{x}_i=(\textbf{x}_F,z_i)$, where $z_i$ indicates the depth of $\partial \textbf{D}_i$
and $\textbf{x}_F$ denotes the horizontal coordinates of a focal point at this depth.
Note that  boundaries $\partial \textbf{D}_0$ and $\partial \textbf{D}_i$ in $2D$ and $3D$ are lines and planes, respectively 
(for a comprehensive analysis of generalized Marchenko concepts in $2D$ and $3D$, see \cite{Wapenaar2018}).
The focusing function $f_1(\textbf{x}_0,\textbf{x}_i,t)$ is the solution of the source-free wave
equation in a truncated medium which focuses at the focal point ${\textbf{x}_i}$.
We define the truncated medium as being identical to the physical medium between $\partial \textbf{D}_i$ and $\partial \textbf{D}_0$, and reflection-free elsewhere (\cite{Wapenaar2014}).
The focusing function $f_1(\textbf{x}_0,\textbf{x}_i,t)$ is decomposed into down- and up-going components, indicated by $f_1^+(\textbf{x}_0,\textbf{x}_i,t)$ 
and $f_1^-(\textbf{x}_0,\textbf{x}_i,t)$, respectively.
The down-going component of the focusing function, $f_1^+(\textbf{x}_{0},\textbf{x}_{i},t)$, is the inverse 
of the transmission response $T(\textbf{x}_i,\textbf{x}_0,t')$ of the above mentioned truncated medium, i.e.
\begin{equation}
 \int_{\partial \textbf{D}_0} d\textbf{x}_{0} \int_0^{\infty} T(\textbf{x}^{''}_i,\textbf{x}_{0},t') f_{1}^+(\textbf{x}_{0},\textbf{x}_{i},t-t')dt' = \delta(\textbf{x}^{''}_{F} -\textbf{x}_{F})\delta(t),
\label{eq:INVERSE_T}
\end{equation}
where $\delta(\textbf{x}^{''}_{F} -\textbf{x}_{F})$ is a two-dimensional delta function along $\partial \textbf{D}_i$.
Both $f_1^+(\textbf{x}_{0},\textbf{x}_{i},t)$ and  $T(\textbf{x}_{i},\textbf{x}_{0},t')$ can be decomposed
into direct and coda components, indicated by $_d$ and $_m$ subscripts, respectively:
\begin{equation}
 f_1^+(\textbf{x}_{0},\textbf{x}_{i},t)  =  f_{1d}^+(\textbf{x}_{0},\textbf{x}_{i},t)+f_{1m}^+(\textbf{x}_{0},\textbf{x}_{i},t)
\label{eq:3}
\end{equation}
and
\begin{equation}
 T(\textbf{x}_{i},\textbf{x}_{0},t)  =  T_d(\textbf{x}_{i},\textbf{x}_{0},t)+T_m(\textbf{x}_{i},\textbf{x}_{0},t).
\label{eq:4}
\end{equation}
Using source-receiver reciprocity, Eq.  \ref{eq:INVERSE_T} can be generalized as:
\begin{equation}
 \int_{\partial \textbf{D}_i} d\textbf{x}_{i} \int_0^{\infty} T_d(\textbf{x}_{i},\textbf{x}^{''}_{0},t') f_{1d}^+(\textbf{x}_{0},\textbf{x}_{i},t-t')dt' = \delta(\textbf{x}^{{''}}_{H} -\textbf{x}_{H})\delta(t),
\label{eq:7}
\end{equation}
where $\delta(\textbf{x}^{''}_{H} -\textbf{x}_{H})$ is now a two-dimensional delta function along $\partial \textbf{D}_0$.
The up-going component of the focusing function, $f_1^-(\textbf{x}^{'}_{0},\textbf{x}_{i},t)$, is by definition the reflection response of the \textit{truncated} medium to $f_1^+(\textbf{x}_{0},\textbf{x}_{i})$,
and it is equivalent to:
\begin{equation}
 f_1^-(\textbf{x}^{'}_{0},\textbf{x}_{i},t) =  \int_{\partial \textbf{D}_0} d\textbf{x}_{0} \int_0^{\infty} R^{}(\textbf{x}^{'}_{0},\textbf{x}_{0},t') f_1^+(\textbf{x}_{0},\textbf{x}_{i},t-t')dt',
\label{eq:5}
\end{equation}
where $R^{}(\textbf{x}^{'}_{0},\textbf{x}_{0},t)$ is the impulse reflection response (with the source ignited at time $t=0$ to allow standard Marchenko derivations) at the
surface of the \textit{physical} medium, with $\textbf{x}^{'}_{0},\textbf{x}_{0}$ denoting receiver/source locations.
This relationship is valid for $ -t_d +\varepsilon < t < t_d + \varepsilon $, where $t_d$ is the one-way traveltime from a surface point ${\textbf{x}}^{'}_{0}$ to $\textbf{x}_i$ and $\varepsilon$ is a small
positive  value accounting for the finite bandwidth of the data. 
Note that, unlike for the original Marchenko scheme, we have chosen an asymmetric time interval, 
following \cite{Zhang2019}. For this time interval, 
the coda of the down-going focusing function, namely  $f_{1m}^+(\textbf{x}^{'}_{0},\textbf{x}_{i},t)$, satisfies the following relationship:
\begin{equation}
 f_{1m}^+(\textbf{x}^{'}_{0},\textbf{x}_{i},t) = \int_{\partial \textbf{D}_0} d\textbf{x}_{0} \int^0_{-\infty} R^{}(\textbf{x}^{'}_{0},\textbf{x}_{0},-t') f_1^-(\textbf{x}_{0},\textbf{x}_{i},t-t')dt'.
\label{eq:6}
\end{equation}
Next we project the focusing functions to the surface. The projected focusing functions $v^-$ and $v^+_m$ are then introduced as:
\begin{equation}
 v^-(\textbf{x}^{'}_{0},\textbf{x}^{''}_{0},t,z_i) = \int_{\partial \textbf{D}_i} d\textbf{x}_{i} \int_0^{\infty} T_d(\textbf{x}_{i},\textbf{x}^{''}_{0},t') f_{1}^-(\textbf{x}^{'}_{0},\textbf{x}_{i},t-t')dt' 
\label{eq:10}
\end{equation}
and
\begin{equation}
 v^+_m(\textbf{x}^{'}_{0},\textbf{x}^{''}_{0},t,z_i) = \int_{\partial \textbf{D}_i} d\textbf{x}_{i} \int_0^{\infty} T_d(\textbf{x}_{i},\textbf{x}^{''}_{0},t') f_{1m}^+(\textbf{x}^{'}_{0},\textbf{x}_{i},t-t')dt',
\label{eq:11}
\end{equation}
where the variable $z_i$ indicates that these functions depend on the depth level along which standard Marchenko focusing functions are defined. 
Note that differently than in previous literature (\cite{van2016adaptive,Zhang2019}) we now make explicit the dependence of $v^-$ and $v^+_m$ on $z_i$ (\cite{10.1093/gji/ggaa005}).
By convolving and integrating in space along $\partial\textbf{D}_i$ both sides of Eqs. \ref{eq:5} and \ref{eq:6}  with $T_d$ as indicated in Eq. \ref{eq:7}, we obtain:
\begin{equation}
 v^-(\textbf{x}^{'}_0,\textbf{x}^{''}_0,t,t_2) =  \int_{\partial \textbf{D}_0} d\textbf{x}_0 \int_0^{\infty} R^{}(\textbf{x}^{'}_0,\textbf{x}_0,t') v_m^+(\textbf{x}_0,\textbf{x}^{''}_0,t-t',t_2) dt'  + R^{}(\textbf{x}^{'}_0,\textbf{x}^{''}_0,t),
\label{eq:8}
\end{equation}
and
\begin{equation}
 v^+_m(\textbf{x}^{'}_0,\textbf{x}^{''}_0,t,t_2) =  \int_{\partial \textbf{D}_0} d\textbf{x}_0 \int^0_{-\infty} R^{}(\textbf{x}^{'}_0,\textbf{x}_0,-t')  v^-(\textbf{x}_0,\textbf{x}^{''}_0,t-t',t_2)  dt',
\label{eq:9}
\end{equation}
for $\varepsilon < t < t_2 + \varepsilon$, where for convenience
we have replaced the dependence on $z_i$ by the new variable $t_2=t_2(\textbf{x}^{'}_0,\textbf{x}^{''}_0,z_i)$  corresponding to the two-way traveltime from a
surface point $\textbf{x}^{''}_0$ to the specular reflection at a (hypothetical) interface at level $z_i$ 
and back to the surface point $\textbf{x}^{'}_0$. 
Different from
previous literature on this subject, we make all the relevant variables in $v^-$ and $v^+_m$ explicit, by considering also $t_2$. 
Note that  for $t<\varepsilon$ and $t>t_2 + \varepsilon$ both $v^-$ and $v^+_m$ are zero, which is why the integrals on the right-hand side are evaluated only for 
the time interval $\epsilon<t<t_2+\epsilon$.
Using the time-domain formalism introduced in \cite{vanderNeutetal2015} we rewrite Eqs. \ref{eq:8} and \ref{eq:9} as:

\begin{equation}
 {v}^{-}(\textbf{x}^{'}_0,\textbf{x}^{''}_0,t,t_2)  = (\Theta_{\varepsilon}^{t_2+\varepsilon} R +\Theta_{\varepsilon}^{t_2+\varepsilon}\textbf{R}v_m^{+})(\textbf{x}^{'}_0,\textbf{x}^{''}_0,t,t_2),   \\
  \label{eq:16}
\end{equation}
and
\begin{equation}
  {v}_m^{+}(\textbf{x}^{'}_0,\textbf{x}^{''}_0,t,t_2) = (\Theta_{\varepsilon}^{t_2+\varepsilon}\textbf{R}^{\star}{v}^{-})(\textbf{x}^{'}_0,\textbf{x}^{''}_0,t,t_2),\\
 \label{eq:16+}
\end{equation}
where $\textbf{R}$ indicates a convolution integral operator of the measured data $R$ with any wavefield, the superscript $\star$ indicates time-reversal and $\Theta_{\varepsilon}^{t_2+\varepsilon}$ is a muting operator removing values outside of the interval $(\varepsilon,t_2+\varepsilon)$. 

Terms in Eq. \ref{eq:16} are rearranged using Eq. \ref{eq:16+} to get:
\begin{equation}
({I} - \Theta_{\varepsilon}^{t_2+\varepsilon} \textbf{R} \Theta_{\varepsilon}^{t_2+\varepsilon} \textbf{R}^{\star} ) v^{-}(\textbf{x}^{'}_0,\textbf{x}^{''}_0,t,t_2)  = \Theta_{\varepsilon}^{t_2+\varepsilon}R(\textbf{x}^{'}_0,\textbf{x}^{''}_0,t),
\label{eq:series}
\end{equation}

which, under standard convergence conditions (\cite{Fokkema2013}), is solved by:

\begin{equation}
{v}^{-}(\textbf{x}^{'}_0,\textbf{x}^{''}_0,t,t_2) = \Theta_{\varepsilon}^{t_2+\varepsilon} R(\textbf{x}^{'}_0,\textbf{x}^{''}_0,t) + \left[ \sum_{M=1}^{\infty} (\Theta_{\varepsilon}^{t_2+\varepsilon} \textbf{R} \Theta_{\varepsilon}^{t_2+\varepsilon} \textbf{R}^{\star})^M \Theta_{\varepsilon}^{t_2+\varepsilon} R \right] (\textbf{x}^{'}_0,\textbf{x}^{''}_0,t).
\label{eq:solution}
\end{equation}

This procedure allows to retrieve ${v}^{-}(\textbf{x}^{'}_0,\textbf{x}^{''}_0,t,t_2)$,  whose last event, when its two-way travel time $t$ is equal to $t_2(x_0',x_0'',z_i)$ is a transmission loss compensated primary
reflection in $R^{}(\textbf{x}^{'}_0,\textbf{x}^{''}_0,t)$ (\cite{Zhang2019}). 
In practice, the transmission loss compensated primary is obtained by computing ${v}^{-}$ via Eq. \ref{eq:solution} for all values $t_2$ (i.e., by considering
the corresponding windowing operator $\Theta_\varepsilon^{t_2+\varepsilon}$), and by storing results in a new, parallel dataset at $t=t_2$. 
Similarly to other Marchenko schemes, in practical applications only a few terms of the series in Eq. \ref{eq:solution}
need to be computed
to achieve proper convergence (\cite{broggini2014data}).
Moreover, following \cite{zhang2018marchenko}, instead of computing $t_2$ as the space- and model-dependent two-way traveltime via a chosen depth level $z_i$, we can evaluate 
Eq. \ref{eq:solution} for all possible \textit{constant} values $\bar{t}_2$ (to include values large enough to allow waves to reach the bottom of the model 
and come back to the surface) and store results at $t=\bar{t}_2$.  
In this way the (transmission-compensated) primary reflection response in $R^{}(\textbf{x}^{'}_0,\textbf{x}^{''}_0,t)$ is then fully retrieved.

\subsection{Extension to horizontal plane-wave data}
\label{horizontal_paragraph}
In this paper, following a similar approach to what  was recently proposed to extend Marchenko redatuming from point-source to horizontal plane-wave concepts (\cite{MelesVirtual}),
we consider integral representations of the projected focusing functions $v^-$ and $v^+_m$.
More precisely, we first define new projected focusing functions $V^-(\textbf{x}^{'}_0,t,t_2)$ and $V^+_m(\textbf{x}^{'}_0,t,t_2)$ as:

%

\begin{equation}
 V^-(\textbf{x}^{'}_0,t,T_2) \equiv \int_{\partial \textbf{D}_0} d\textbf{x}^{''}_0 v^-(\textbf{x}^{'}_0,\textbf{x}^{''}_0,t,t_2),
\label{eq:15}
\end{equation}
and
\begin{equation}
 V^+_m(\textbf{x}^{'}_0,t,T_2) \equiv  \int_{\partial \textbf{D}_0} d\textbf{x}^{''}_0 v^+_m(\textbf{x}^{'}_0,\textbf{x}^{''}_0,t,t_2),
\label{eq:14}
\end{equation}
where $T_2=T_2(\textbf{x}^{'}_0,z_i)$ is 
the two-way traveltime of a horizontal plane-wave  propagating
down from the surface to the specular reflection at a (hypothetical) interface at level $z_i$ 
and back to the surface point $\textbf{x}^{'}_0$.
We then integrate Eqs. \ref{eq:8} and \ref{eq:9} along $\partial \textbf{D}_0$ to obtain:
%
\begin{equation}
 V^-(\textbf{x}^{'}_0,t,T_2) = \int_{\partial \textbf{D}_0} d\textbf{x}_0 \int_0^{\infty} R^{}(\textbf{x}^{'}_0,\textbf{x}_0,t') V_m^+(\textbf{x}_0,t-t',T_2) dt'  + R_{\rm PW} ^{}(\textbf{x}^{'}_0,t),
\label{eq:12}
\end{equation}
and
\begin{equation}
 V^+_m(\textbf{x}^{'}_0,t,T_2) =  \int_{\partial \textbf{D}_0} d\textbf{x}_0 \int_{-\infty}^{0} R^{}(\textbf{x}^{'}_0,\textbf{x}_0,-t')  V^-(\textbf{x}_0,t-t',T_2)  dt',
\label{eq:13}
\end{equation}
for $\varepsilon < t< T_2 + \varepsilon$ and 
where $R_{\rm PW}^{}(\textbf{x}^{'}_0,t) \equiv \int_{\partial \textbf{D}_0} d\textbf{x}^{''}_0 R^{}(\textbf{x}^{'}_0,\textbf{x}^{''}_0,t)$ is by definition the horizontal plane-wave source response 
of the medium (i.e., the source emits a vertically downward propagating plane wave).
Using again the time-domain formalism we can therefore rewrite Eqs. \ref{eq:12} and \ref{eq:13} as:
%
\begin{equation}
\begin{aligned}
 {V}^{-}(\textbf{x}^{'}_0,t,T_2)  = (\Theta_{\varepsilon}^{T_2+\varepsilon} R_{\rm PW} + \Theta_{\varepsilon}^{T_2+\varepsilon}\textbf{R}V_m^{+}) (\textbf{x}^{'}_0,t,T_2),  \\
  \label{eq:l16}
\end{aligned}
\end{equation}
and
\begin{equation}
\begin{aligned}
 {V}_m^{+}(\textbf{x}^{'}_0,t,T_2) = (\Theta_{\varepsilon}^{T_2+\varepsilon}\textbf{R}^{\star}{V}^{-})(\textbf{x}^{'}_0,t,T_2),\\
 \label{eq:l16+}
\end{aligned}
\end{equation}
and therefore:
%
%
%
\begin{equation}
({I} - \Theta_{\varepsilon}^{T_2+\varepsilon} \textbf{R} \Theta_{\varepsilon}^{T_2+\varepsilon} \textbf{R}^{\star} ){V}^{-}(\textbf{x}^{'}_0,t,T_2)  = \Theta_{\varepsilon}^{T_2+\varepsilon} R_{\rm PW}(\textbf{x}^{'}_0,t),
\label{eq:lseries}
\end{equation}
which is solved by:
\begin{equation}
{V}^{-}(\textbf{x}^{'}_0,t,T_2) = \Theta_{\varepsilon}^{T_2+\varepsilon} R_{\rm PW}(\textbf{x}^{'}_0,t) + \left[ \sum_{M=1}^{\infty} (\Theta_{\varepsilon}^{T_2+\varepsilon} \textbf{R} \Theta_{\varepsilon}^{T_2+\varepsilon} \textbf{R}^{\star})^M \Theta_{\varepsilon}^{T_2+\varepsilon} {R_{\rm PW}} \right] (\textbf{x}^{'}_0,t).
\label{eq:1solution}
\end{equation}
This procedure allows to retrieve ${V}^{-}(\textbf{x}^{'}_0,t,T_2)$,  whose last event, when its two-way travel time $t$ is equal to $T_2(x_0',z_i)$, is a transmission loss compensated primary
reflection in $R_{\rm PW}^{}(\textbf{x}^{'}_0,t)$. 
Instead of computing $T_2$ as the space- and model-dependent two-way traveltime via a chosen depth level $z_i$, we can evaluate 
Eq. \ref{eq:1solution} for \textit{constant} values $\bar{T}_2$.  
By computing Eq. \ref{eq:1solution} for all possible constant values $\bar{T}_2$ and storing results at $t=\bar{T}_2$, the (transmission-compensated) primary reflection
response in $R_{\rm PW}^{}(\textbf{x}^{'}_0,t)$ is then fully retrieved. Note that in practical applications, the integrals along $\partial \textbf{D}_0$ in Eqs.  \ref{eq:15}-\ref{eq:13} and 
in the definition of $R_{PW}$ are replaced by summations over source locations.

\subsection{Extension to dipping plane-wave data}
\label{dipping_paragraph}
\begin{figure}
   \centering
   \includegraphics[width=0.99\textwidth]{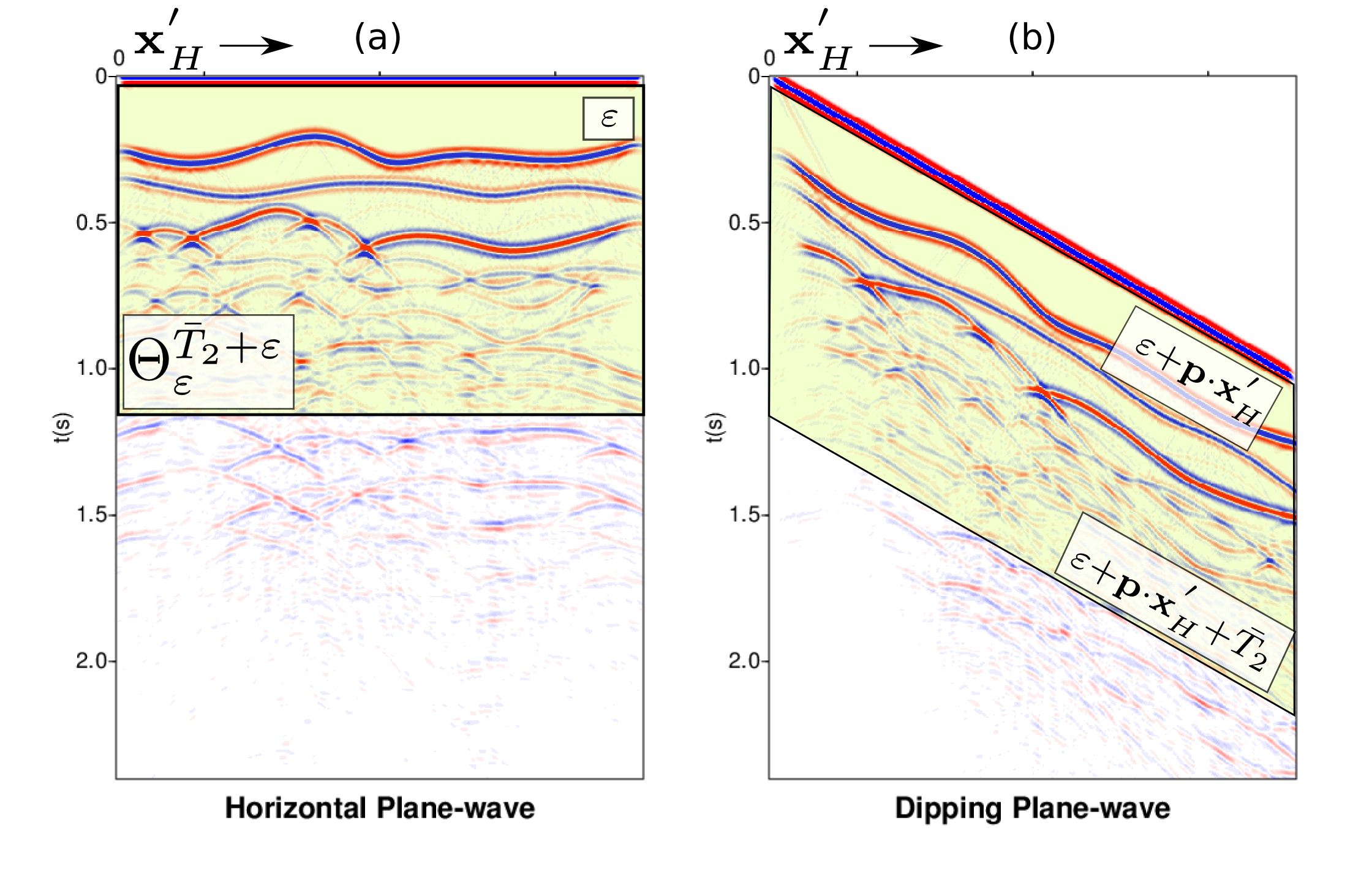}
   \caption{The shaded green areas show the support of representative muting operators for horizontal ($\Theta_{\varepsilon}^{\bar{T}_2+\varepsilon}$ in (a)) 
   and dipping ($\Theta_{\varepsilon+\textbf{p}\cdot \textbf{x}_H^{'}}^{\varepsilon+\bar{T}_2+\textbf{p}\cdot \textbf{x}_H^{'}}$ in (b)) plane-wave sources (the corresponding data are shown in the background).}
   \label{Fig:1.png}
    \end{figure}

In standard Marchenko derivations it is assumed that point sources are fired at $t=0$ (\cite{Wapenaar2014,Zhang2019}). 
Since dipping plane-waves are 
associated with \textit{many} sources excited at \textit{different} times, we cannot expect standard algorithms, such as that in Eq. \ref{eq:1solution},
to predict primaries when delayed source gathers are considered.
To illustrate how to proceed when dipping plane-waves are taken into account, we first consider the obvious corresponding projected focusing functions:
\begin{equation}
V^-(\textbf{x}_0^{'},\textbf{p},t,T_2)\equiv\int_{\partial \textbf{D}_0} d\textbf{x}^{''}_0 v^-(\textbf{x}^{'}_0,\textbf{x}^{''}_0,t-\textbf{p}\cdot \textbf{x}^{''}_H,{t}_2)
\label{eq:delayed1}
\end{equation}
and
\begin{equation}
V^+_m(\textbf{x}_0^{'},\textbf{p},t,T_2)\equiv\int_{\partial \textbf{D}_0} d\textbf{x}^{''}_0 v^+_m(\textbf{x}^{'}_0,\textbf{x}^{''}_0,t-\textbf{p}\cdot \textbf{x}^{''}_H,{t}_2)
\label{eq:delayed2}
\end{equation}
where $\textbf{p}$ is a  ray parameter vector and $T_2=T_2(\textbf{x}^{'}_0,\textbf{p},z_i)$ is the two-way traveltime of a plane-wave with ray parameter $\textbf{p}$, propagating
down from the surface to the specular reflection at a (hypothetical) interface at level $z_i$ 
and back to the surface point $\textbf{x}^{'}_0$.
Substituting Eqs. \ref{eq:8} and \ref{eq:9} into Eqs. \ref{eq:delayed1} and \ref{eq:delayed2},
and indicating the reflection response associated
with a dipping plane-wave source characterized by ray parameter vector $\textbf{p}$ as 
$R_{DW}(\textbf{x}^{'}_0,\textbf{p},t)\equiv \int_{\partial \textbf{D}_0} d\textbf{x}_0^{''} R(\textbf{x}^{'}_0,\textbf{x}^{''}_0,t-\textbf{p}\cdot \textbf{x}^{''}_H)$,
we obtain:
\begin{equation}
 V^-(\textbf{x}_0^{'},\textbf{p},t,T_2) = \int_{\partial \textbf{D}_0} d\textbf{x}_0 \int_0^{\infty} R^{}(\textbf{x}^{'}_0,\textbf{x}_0,t') V_m^+(\textbf{x}_0,\textbf{p},t-t',T_2) dt'  + R_{DW}(\textbf{x}^{'}_0,\textbf{p},t),
\label{eq:S12}
\end{equation}
and
\begin{equation}
 V^+_m(\textbf{x}_0^{'},\textbf{p},t,T_2) =  \int_{\partial \textbf{D}_0} d\textbf{x}_0 \int_{-\infty}^{0} R^{}(\textbf{x}^{'}_0,\textbf{x}_0,-t')  V^-(\textbf{x}_0,\textbf{p},t-t',T_2)  dt',
\label{eq:S13}
\end{equation}
for $ \varepsilon +  \textbf{p}\cdot \textbf{x}'_H < t < T_2 + \varepsilon$. 
The relationship between ${V}^{-}(\textbf{x}^{'}_0,\textbf{p},t,T_2)$ and ${V}_m^{+}(\textbf{x}^{'}_0,\textbf{p},t,T_2)$, using again the time-domain formalism, is then established by:
\begin{equation}
\begin{aligned}
 {V}^{-}(\textbf{x}^{'}_0,\textbf{p},t,T_2)  = (\Theta_{\varepsilon+\textbf{p}\cdot \textbf{x}^{'}_H}^{T_2 + \varepsilon} R_{DW} + \Theta_{\varepsilon+\textbf{p}\cdot \textbf{x}^{'}_H}^{T_2 + \varepsilon}\textbf{R}V_m^{+}) (\textbf{x}^{'}_0,\textbf{p},t,T_2),  \\
  \label{eq:Sl16}
\end{aligned}
\end{equation}
and
\begin{equation}
\begin{aligned}
 {V}_m^{+}(\textbf{x}^{'}_0,\textbf{p},t,T_2) = (\Theta_{\varepsilon+\textbf{p}\cdot \textbf{x}^{'}_H}^{T_2 + \varepsilon}\textbf{R}^{\star}{V}^{-})(\textbf{x}^{'}_0,\textbf{p},t,T_2).\\
 \label{eq:Sl16+}
\end{aligned}
\end{equation}
Combining Eqs. \ref{eq:Sl16} and \ref{eq:Sl16+} together we finally get:
\begin{equation}
({I} -\Theta_{\varepsilon+\textbf{p}\cdot \textbf{x}_H'}^{T_2 + \varepsilon} \textbf{R} \Theta_{\varepsilon+\textbf{p}\cdot\textbf{x}_H'}^{T_2 + \varepsilon} \textbf{R}^{\star} ){V}^\textbf{-}(\textbf{x}^{'}_0,\textbf{p},t,T_2)  = \Theta_{\varepsilon+\textbf{p}\cdot \textbf{x}_H'}^{T_2 + \varepsilon} R_{\rm DW}(\textbf{x}^{'}_0,\textbf{p},t),
\label{eq:Sseries}
\end{equation}
which is solved by:
\begin{equation}
\begin{aligned}
{V}^{-}(\textbf{x}^{'}_0,\textbf{p},t,T_2) = \Theta_{\varepsilon+\textbf{p}\cdot \textbf{x}_H'}^{T_2 + \varepsilon} R_{DW}(\textbf{x}^{'}_0,\textbf{p},t) + \left[ \sum_{M=1}^{\infty} (\Theta_{\varepsilon+\textbf{p} \cdot \textbf{x}_H'}^{T_2 + \varepsilon} \textbf{R} \Theta_{\varepsilon+\textbf{p}\cdot \textbf{x}_H'}^{T_2 + \varepsilon} \textbf{R}^{\star})^M \Theta_{\varepsilon+\textbf{p}\cdot \textbf{x}_H'}^{T_2 + \varepsilon} {R_{\rm DW}} \right] (\textbf{x}^{'}_0,\textbf{p},t).
\label{eq:S1solution}
\end{aligned}
\end{equation}
This procedure allows to retrieve $V^{-}(\textbf{x}^{'}_0,\textbf{p},t,T_2)$, 
whose last event, when its two-way travel time $t$ is equal to $T_2(x_0',p,z_i)$, is a transmission loss compensated primary
reflection in $R_{DW}(\textbf{x}^{'}_0,\textbf{p},t)$. 
Note that, in principle,
the muting operators in Eq. \ref{eq:S1solution}, similarly to those in Eqs. \ref{eq:solution} and \ref{eq:1solution}, 
are space- and model-dependent. 
However, in analogy to the previous cases,  
the upper boundary of the muting operators in Eq.  \ref{eq:S1solution} can be taken parallel to the lower one (see Fig. \ref{Fig:1.png}),
thus exhibiting a space-dependent but model-independent shape, i.e. $T_2(x_0',p,z_i) + \varepsilon \approx\varepsilon + \bar{T}_2 + \textbf{p} \cdot \textbf{x}_H^{'}$ for a generic \textit{constant} value $\bar{T}_2$.
By computing Eq. \ref{eq:S1solution} for all possible constant values $\bar{T}_2$ and 
storing results at $t=\bar{T}_2 + \textbf{p} \cdot  \textbf{x}_H^{'}$, 
the (transmission-compensated) primary reflection
response in $R_{\rm DW}^{}(\textbf{x}^{'}_0,\textbf{p},t)$ is then fully retrieved.
The performance of the algorithm in Eq. \ref{eq:S1solution} is assessed in the following numerical examples.
\section{Numerical Examples}
We explore the potential of the proposed scheme for the retrieval of plane-wave source primary reflections with numerical examples involving increasingly complex 2D models.
Evaluation of the series in Eq. \ref{eq:1solution} requires computation of the operators
$\textbf{R}$ and $\textbf{R}^{\star}$ and of the plane-wave reflection response $R_{\rm PW}(\textbf{x}^{'}_0,t)$. 
The reflection responses in $\textbf{R}$ and $\textbf{R}^{\star}$ need to be recorded with wide band and  properly sampled (according to Nyquist criterion in space and time) co-located sources and
receivers placed at the surface of the model. In the following numerical examples, source-receiver sampling is
set to $10$m, while gathers $R_{\rm PW}(\textbf{x}^{'}_0,t)$ are computed with a $20$ Hz Ricker Wavelet. All data used here are simulated with a Finite Difference 
Time Domain solver (\cite{thorbecke2017implementation}). 
    \begin{figure} 
  \centering
   \includegraphics[width=0.99\textwidth]{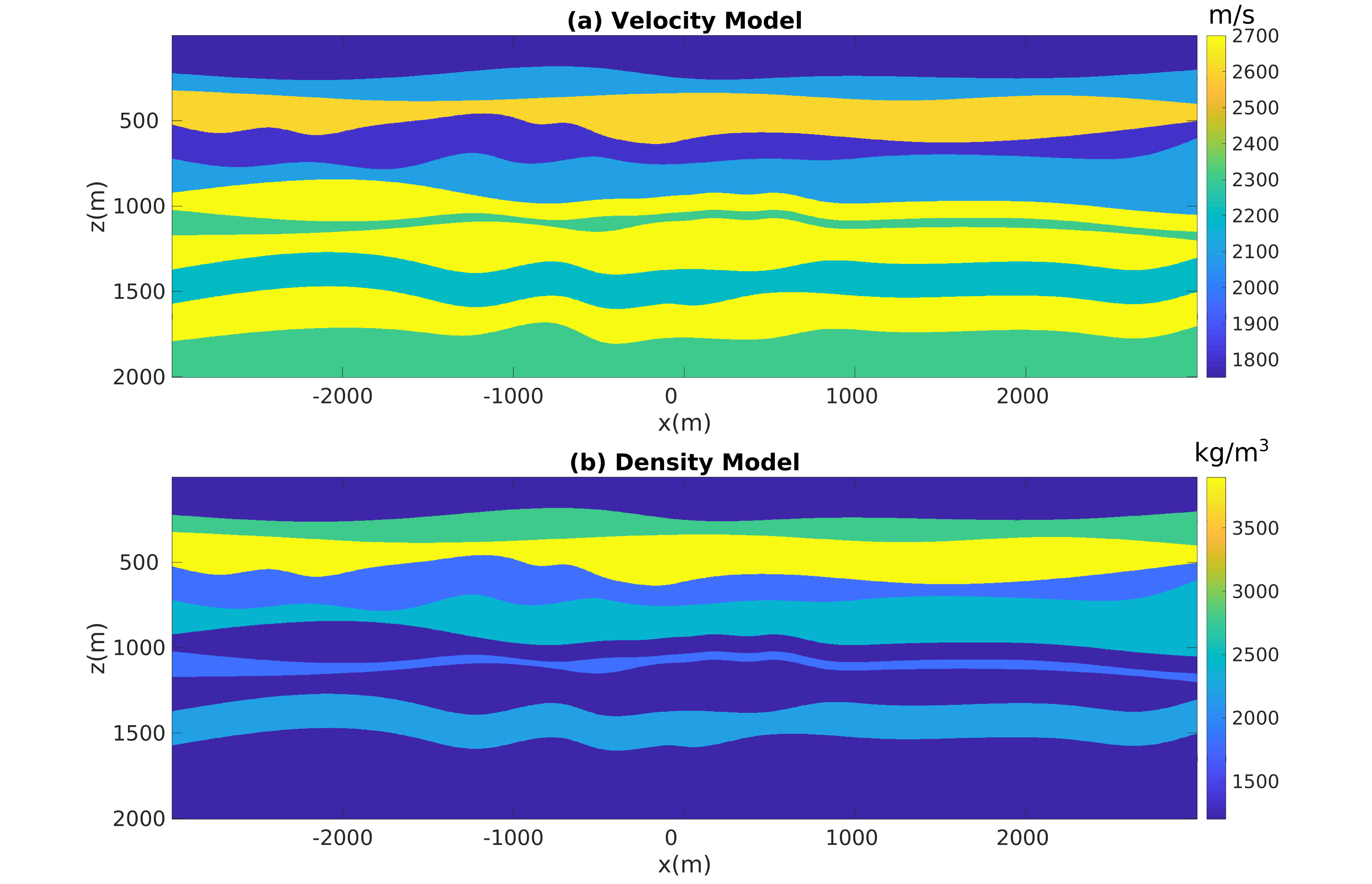}
   \caption{(a) Velocity and (b) density models used in the first numerical experiment.
}
  \label{01_model}
  \end{figure}

   \begin{figure} 
  \centering
   \includegraphics[width=0.99\textwidth]{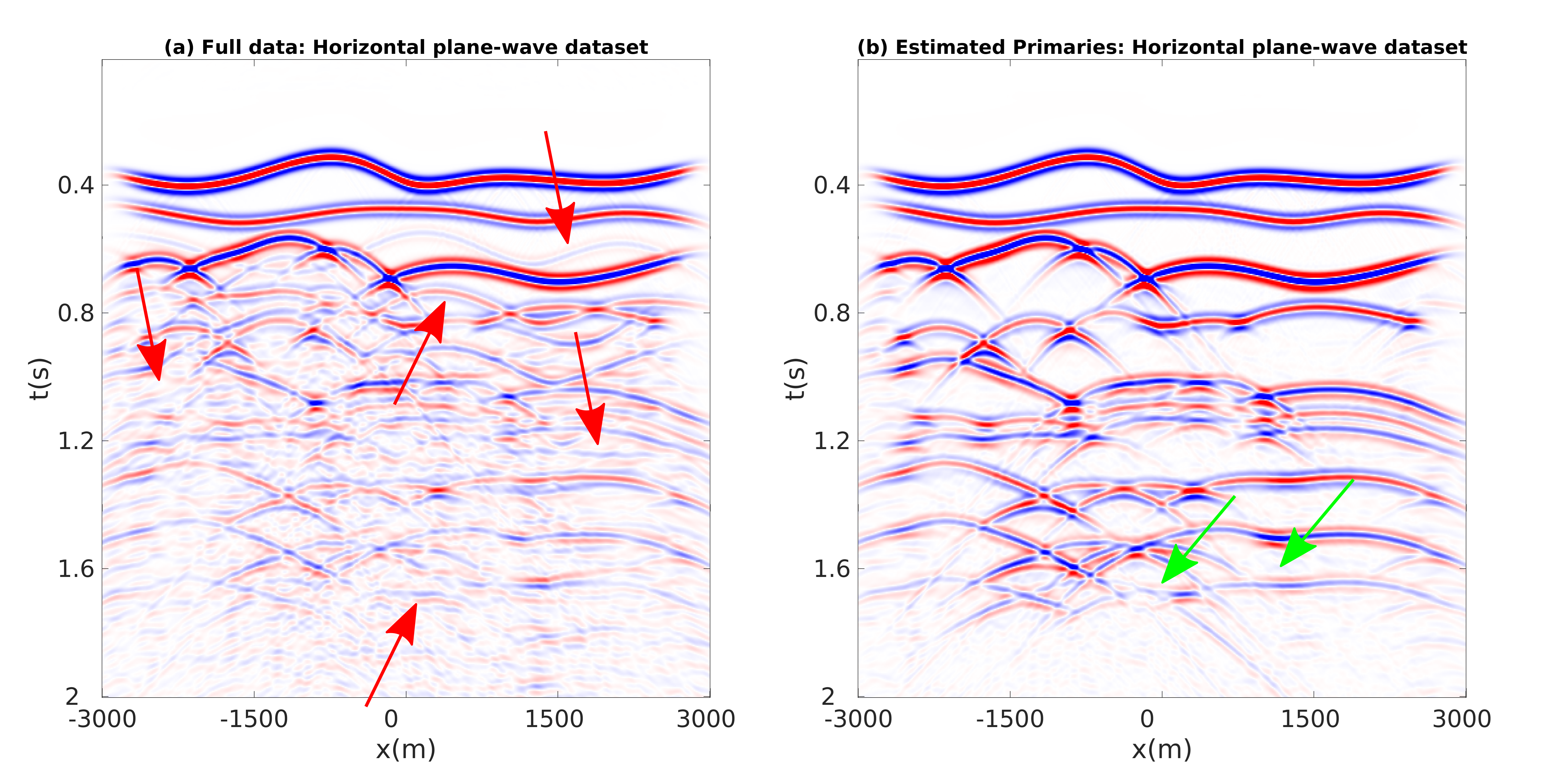} 
   \caption{(a) Full dataset associated with a plane-wave source fired at the surface of the model. Red arrows point at internal multiples. 
   (b) Estimated primaries obtained by computing $\textbf{V}^{-}$ via Eq. \ref{eq:1solution} for all possible values $\bar{T}_2$ and storing results at $t=\bar{T}_2$.
    Differences in amplitude between gathers in (a) and (b) are due to multiple removal and transmission loss compensation.}
  \label{01_data}
  \end{figure}
  
     \begin{figure} 
  \centering
   \includegraphics[width=0.99\textwidth]{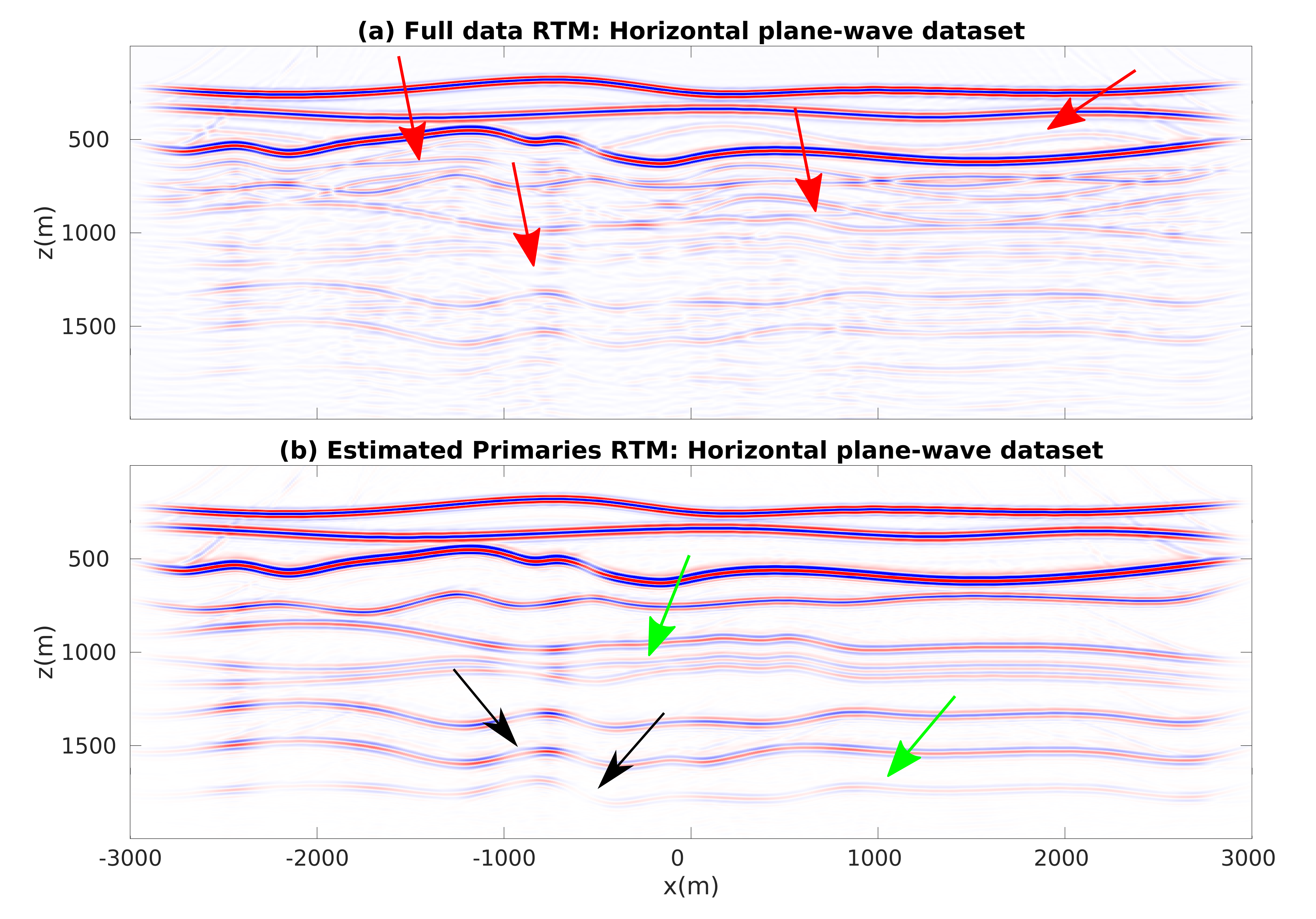} 
   \caption{(a) Standard plane-wave reverse time migration of the dataset in Fig. \ref{01_data}(a). Red arrows point at artefacts related to internal multiples. (b) Standard plane-wave reverse time migration
   of the dataset in Fig. \ref{01_data}(b). Green arrows point at  well resolved interfaces barely visible in (a) due to the superposition of internal multiples. Black arrows point at dipping interfaces
   only partially visible via horizontal plane-wave illumination. Differences in amplitude between images in (a) and (b) are due to multiple removal and transmission loss compensation.}
  \label{01_rtm}
  \end{figure}
  
  For our first numerical experiment we consider a 2D model with gently dipping interfaces (see Fig. \ref{01_model}). The recording surface is reflection free. The dataset associated with a horizontal plane-wave source fired at the surface
  of this model is shown in Fig. \ref{01_data}(a). Notwithstanding the geometrical simplicity of the model, due to the strong impedance variations
  the data are contaminated with many internal multiples, as indicated by the red arrows. We then apply to this dataset the method as described in section \ref{horizontal_paragraph}.
  More precisely, we compute ${V}^{-}$ via Eq. \ref{eq:1solution} for all values $\bar{T}_2$, and by storing results at  $t=\bar{T}_2$ we build a parallel dataset, which theoretically only involves
  primaries. Note that the algorithm is fully data driven, and no model information or any human intervention
  (e.g., picking) is involved in the process. For this dataset we only computed the first $20$ terms of the series in Eq. \ref{eq:1solution}. 
  The result of the procedure is shown in Fig. \ref{01_data}(b). We then image both datasets in Fig. \ref{01_data} via standard plane-wave reverse time migration 
  (based on the zero lag of the cross-correlation between the source and receiver wavefields, \cite{claerbout1985imaging}) using a smoothed version of
  the true velocity distribution in Fig. \ref{01_model} and constant density. Migration results are shown in Fig. \ref{01_rtm}. When the full dataset is migrated, 
  internal multiples contaminate the image as shown in Fig. \ref{01_rtm}(a), producing many false positive artefacts (indicated by red arrows). 
  The image is much cleaner when the dataset in \ref{01_data}(b) is migrated. 
  Each interface is properly recovered, as demonstrated by a comparison between Figs. \ref{01_model} and \ref{01_rtm}(b). 
  Green arrows in \ref{01_rtm}(b) point at physical interfaces which are invisible in Fig. \ref{01_rtm}(a), where they are attenuated by interfering multiple-related artefacts.
  Black arrows point at physical interfaces only partially resolved. The relatively poor performances in imaging dipping interfaces is not due to residual internal multiples, 
  but to the intrinsic limitations of horizontal plane-wave imaging.
  However, note that only one demultipled plane-wave response and a single migration were required to produce the multiple-free image in Fig. \ref{01_rtm}(b).
  We conclude that for gently dipping models horizontal plane-wave datasets are sufficient to produce satisfactory results.
  
  \begin{figure} 
  \centering
   \includegraphics[width=0.99\textwidth]{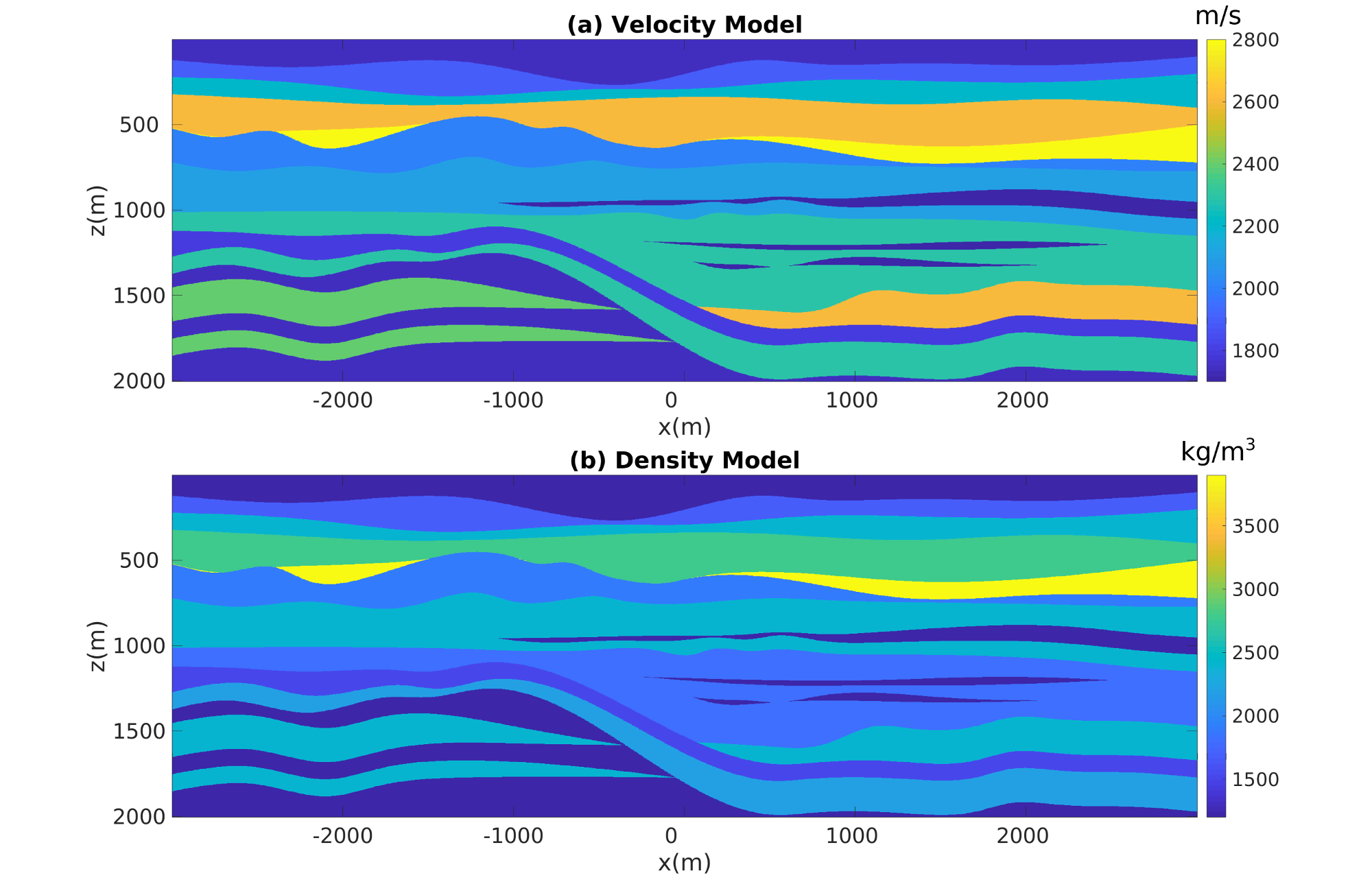}
   \caption{(a) Velocity and (b) density models used in the second numerical experiment.
}
  \label{02_model}
  \end{figure}

   \begin{figure} 
  \centering
   \includegraphics[width=0.94\textwidth]{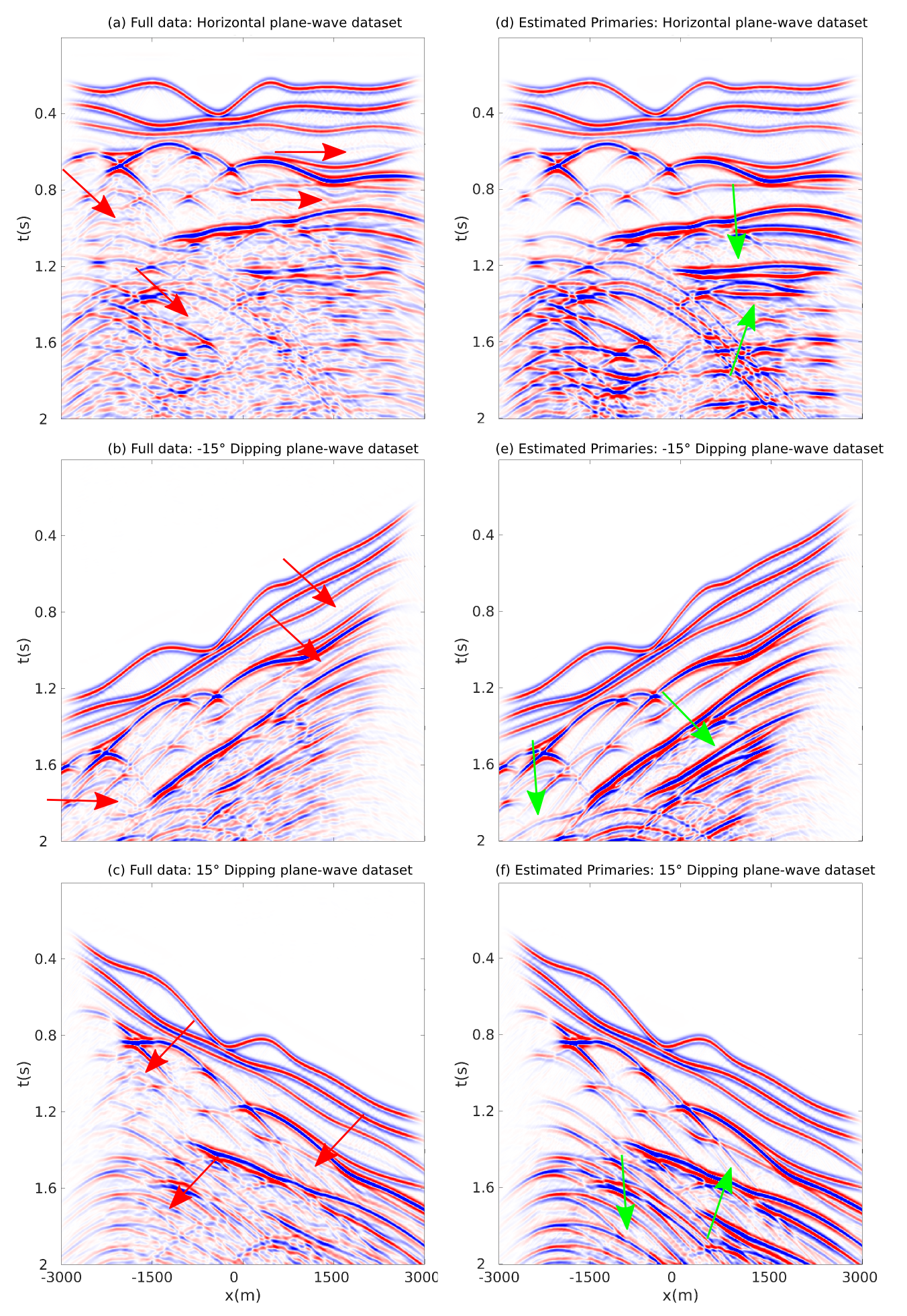} 
   \caption{(a-c): reflection responses associated with plane wave sources at $-15^\circ$, $0^\circ$ and $15^\circ$, respectively.
   Red arrows show internal multiples. (d-f): estimated primaries associated with plane wave sources at $-15^\circ$, $0^\circ$ and $15^\circ$, respectively. 
   Differences in amplitude between gathers in (a-c) and (d-f) are due to multiple removal and transmission loss compensation. Green arrows show
   primaries barely visible in the corresponding full datasets (a-c).}
  \label{02_data}
  \end{figure}
  
     \begin{figure} 
  \centering
   \includegraphics[width=0.99\textwidth]{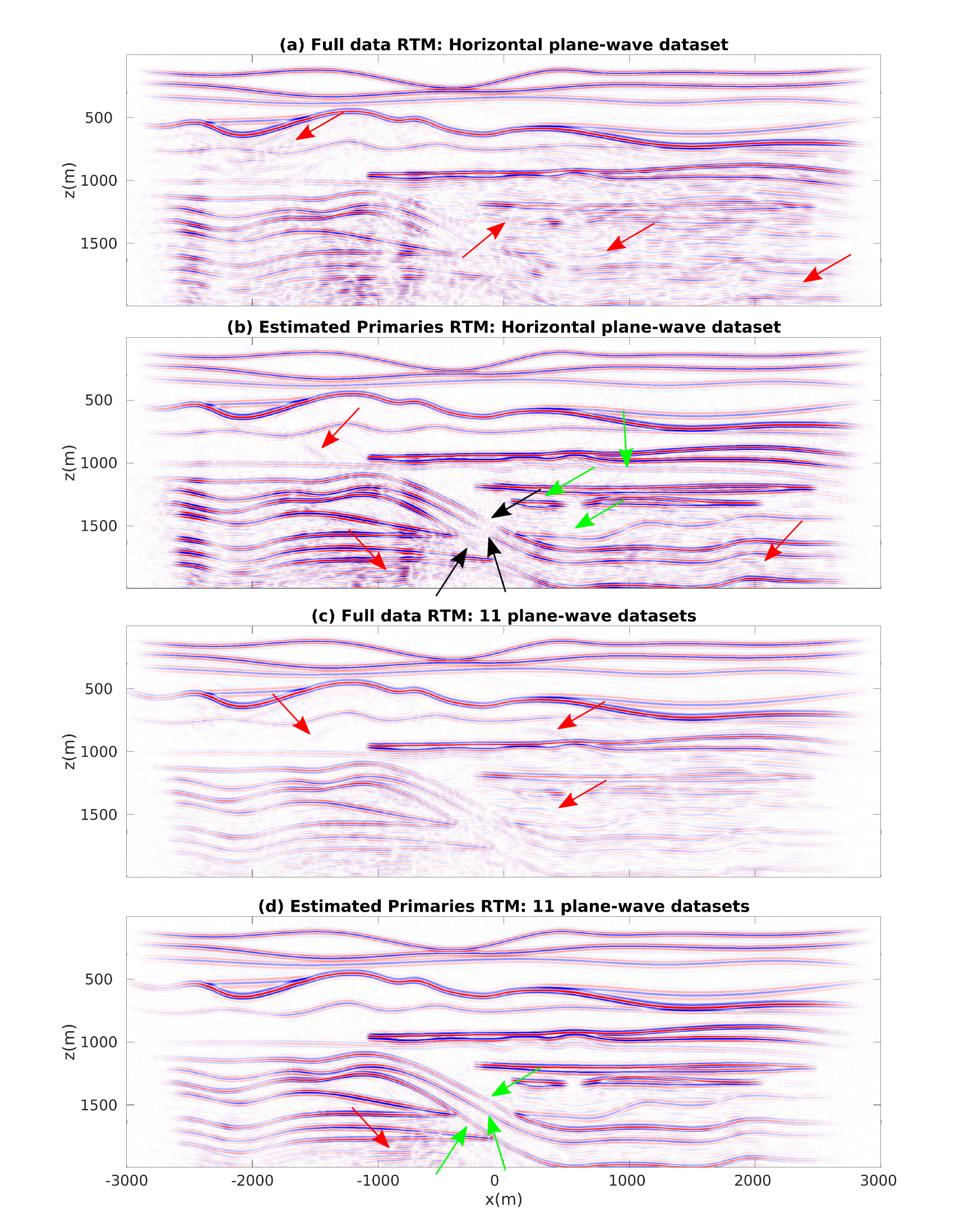} 
   \caption{(a) Standard plane-wave reverse time migration of the dataset in Fig. \ref{02_data}(a). Red arrows point at artefacts related to internal multiples. (b) Standard plane-wave reverse time migration
   of the dataset in Fig. \ref{02_data}(b). Black arrows indicate dipping interfaces that are only partially recovered due to the poor illumination provided by a single plane-wave experiment. Note that
   these interfaces are also not properly imaged in (a). (c) Aggregate RTM of 11 plane wave full datasets (uniformly ranging from $-25^\circ$ to $25^\circ$). Red arrows point at artefacts related 
   to internal multiples. 
   (d) Aggregate Standard RTM of synthesized primaries. 
   Green and red arrows indicate interfaces barely visible in (a) and minor residual artefacts, respectively. 
    Differences in amplitude between images in (a,c) and (b,d) are due to multiples removal and transmission loss compensation (see Fig. \ref{02_data}).}
  \label{02_rtm}
  \end{figure}

In the second example (Fig. \ref{02_model}) we consider a more challenging model 
with critical features for any Marchenko method, i.e. the presence of thin layers, diffractors and dipping layers  (\cite{Wapenaar2014,Zhang2019,dukalski2019handling}). 
We initially follow the same imaging strategy as for the first example.
We first compute the dataset associated with a horizontal plane-wave source fired at the surface of the model shown in Fig. \ref{02_model}(a). 
Also for this dataset we only computed the first $20$ terms of the series in Eq. \ref{eq:1solution}.
Given the complexity of the model, many events, primaries as well as internal multiples
(red arrows) cross each-other, especially in the lower part of the plane-wave gather. Picking specific events in the gather in Fig. \ref{02_data}(a) would be challenging. 
However, as discussed above, our method does not involve any human intervention,
and by applying the same scheme as for the first model we retrieve the dataset shown in Fig. \ref{02_data}(b), where primaries otherwise overshadowed by interfering
multiples are clearly visible (green arrows).
We then migrate datasets in Fig. \ref{02_data}(a) and (b), and show in Fig. \ref{02_rtm}(a) and (b) the corresponding images. 
Large portions of the image in Fig. \ref{02_rtm}(a) associated with the dataset in Fig. \ref{02_data}(a)
are dominated by noise due to the presence of internal multiples (red arrows). 
On the other hand, the image in Fig. \ref{02_rtm}(b), which is associated with the estimated primaries in Fig. \ref{02_data}(d), is much cleaner, with fewer artefacts (red arrows) contaminating limited
domains of the image. Note that relatively poor imaging performances of  dipping interfaces (black arrows in Fig. \ref{02_rtm}(b)) 
are not necessarily associated with shortcomings of the discussed demultiple method but with the intrinsic limitation of what can be illuminated by a single plane-wave experiment.
For this specific model we then decide to process and migrate also dipping plane-wave data. 
In total we then consider 10 additional datasets, uniformly ranging from $-25^\circ$ to $25^\circ$ 
(as discussed in section \ref{dipping_paragraph}, the angle  of the plane-wave is implemented by adding time delays to the shot positions on the horizontal array).
Representative dipping plane-wave data are shown in Fig. \ref{02_data}(b,c), next to the corresponding processed gathers (in Fig. \ref{02_data}(e,f)). Red and green arrows
point again at internal multiples and recovered primaries, respectively. We finally consider aggregate plane-wave migrated images. By migrating a total of $11$ full-data gathers,
the image in Fig. \ref{02_rtm}(c) is obtained. While thanks to the better illumination the improvement over the image in Fig. \ref{02_rtm}(a) is clear, 
some of the key features of the final result are still misleading (red arrows point at false positives associated with the migration of internal multiples).
A significantly better result is obtained when the $11$ processed gathers are imaged and stacked (Fig. \ref{02_rtm}(d)). 
The dipping features poorly visible in (b) are now properly resolved.
This example shows that the proposed method can successfully process dipping plane-wave datasets and therefore
benefit from the corresponding improved illumination. Residual artefacts in the migrated image indicated by the red arrow in Fig. \ref{02_rtm}(d) are likely
due to the presence of thin layers, diffractors and dipping layers that are known to be critical in Marchenko methods.

\section{Discussion}
\begin{figure} 
  \centering
   \includegraphics[width=0.99\textwidth]{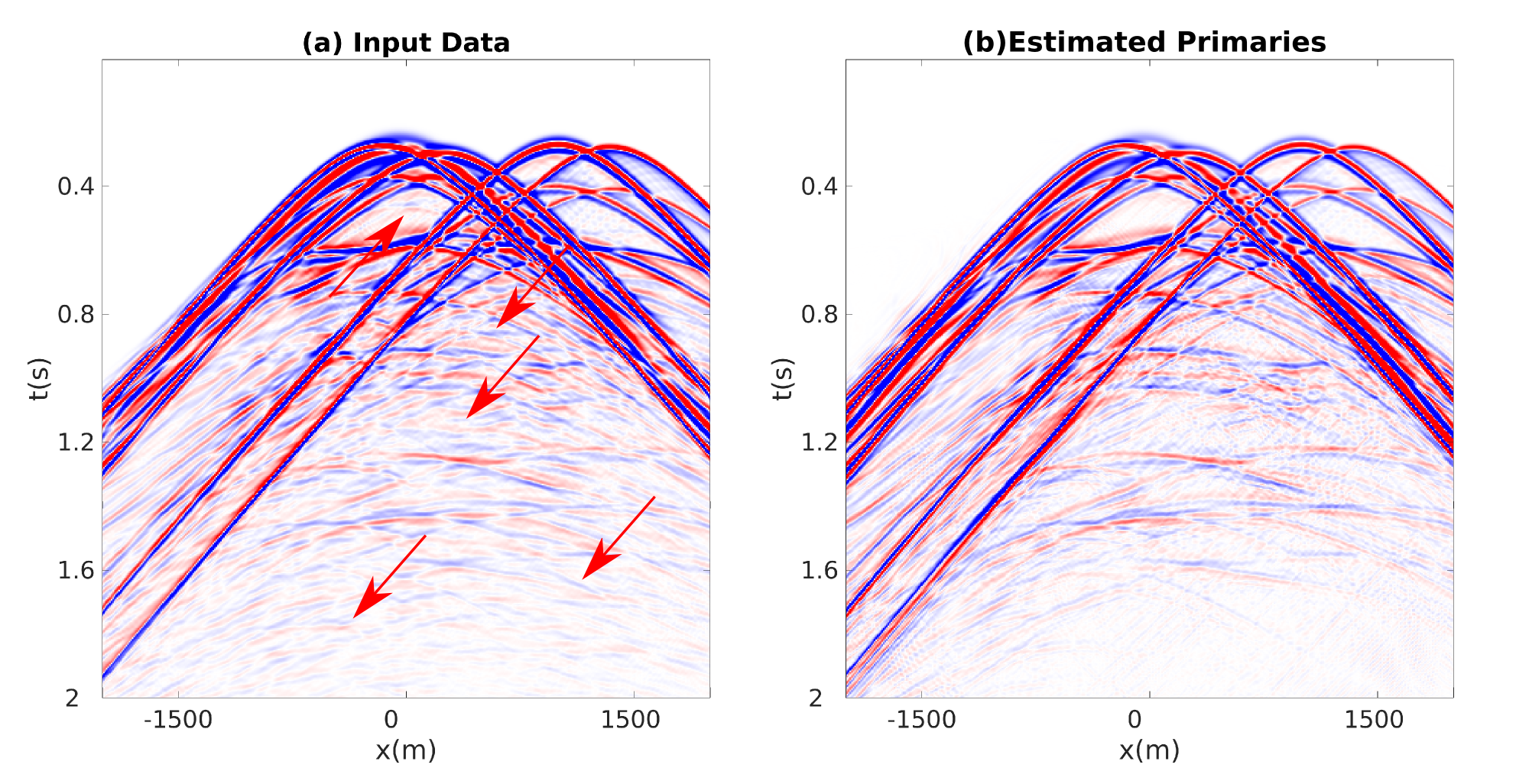} 
   \caption{(a) Full dataset associated with 5 point sources with different spectrum content fired at the same time. Red arrows point at internal multiples. 
   (b) Corresponding estimated primaries.}
  \label{blended}
  \end{figure}

In section \ref{horizontal_paragraph} we have extended a recently proposed primary synthesis method, devised for point source gathers, to horizontal plane-wave source data. 
The new scheme still needs full point-source data as input, but its output is a horizontal plane-wave response.
The method is based on integration of point-source responses over the acquisition surface (e.g., Eqs. \ref{eq:8} and \ref{eq:9}), which allows the derivation of
relationships associated with plane-wave sources (e.g., Eqs. \ref{eq:12} and \ref{eq:13}).
Both the point-source and plane-wave primary synthesis methods are totally data driven, and both are implemented by inversion of the same family of linear operators, i.e.:
\begin{equation}
{I} - \Theta_{\varepsilon}^{\bar{T}_2+\varepsilon} \textbf{R} \Theta_{\varepsilon}^{\bar{T}_2+\varepsilon} \textbf{R}^{\star} 
\label{eq:operators}
\end{equation}
Each operator is defined by a different value of $\bar{T}_2$.
In previous literature that underlies this contribution, an integration over the focusing surface was used to adapt Greens' 
functions redatuming methods to virtual plane-wave redatuming (\cite{MelesVirtual}).
While conceptually similar, there is a subtle yet very important difference between the
methods discussed here and previous methods on virtual plane-wave redatuming.
Whereas in any Marchenko redatuming scheme (e.g. for point or plane virtual sources) a  \textit{different}, model dependent,
window operator for \textit{each} point or plane is required, as focusing is achieved in the subsurface,
the window operators discussed here are the \textit{same} for \textit{each} input data, as the focusing operators are projected to the surface.
Since the operators in Eq. \ref{eq:operators}  are linear and do not depend on the specific gather they are applied to,
any linear combination of point-source data can be processed at once, provided that all the corresponding sources are fired at the same time (see section \ref{horizontal_paragraph} for more details).
The proposed method can then be used, without any modification, to blended-source data as well as to individual point sources and 
horizontal plane-wave gathers. This is shown in Fig. \ref{blended}, where the algorithm is applied
to a dataset associated with 5 sources with different spectra fired at the same time (Fig. \ref{blended}(a)). 
Application of the proposed scheme results in the gather shown in Fig. \ref{blended}(b). A nearly identical result 
(relative difference smaller than $0.1\%$) is achieved when the method is applied to each single point source gather
separately,  after which the corresponding results are summed together.\\
In section \ref{dipping_paragraph} we extended the primary synthesis method for dipping plane-wave source data, which helps to improve the illumination of dipping interfaces in the subsurface.

\section{Conclusions}

We have shown that recent advances in data domain Marchenko methods can be extended to incorporate plane-wave source concepts.
More specifically, we have discussed how to retrieve estimates of the primary responses to a plane-wave source.
The retrieved primaries can then be used via standard reverse time migration to produce images free of artefacts
related to internal multiples.
Whereas previous data domain Marchenko methods are applied to point source gathers and therefore tend to be rather expensive for
large datasets, the proposed method  provides  good imaging results by only involving a small number of primary synthesis steps and the corresponding plane-wave reverse time migration.
The plane-wave source primary synthesis algorithm discussed in this paper could then be used as an initial and inexpensive processing step,
potentially guiding more expensive target imaging techniques. 
In this paper we have only discussed 2D examples and internal multiples, but an obvious extension would be allowing surface source primary synthesis in 3D problems as well as incorporating
free surface multiples. Finally, applications of data domain Marchenko methods to field data have already been performed. Future work will then
focus on applying plane-wave primary synthesis methods to field data too.
\section*{Acknowledgments}
The authors thank Max Holicki (Delft University of Technology)
for his collaboration and fruitful discussions. 
This work is partly funded by the European Research Council (ERC) under the European Union's Horizon 2020 research and innovation programme (grant agreement No: 742703).
 
\section*{Data and materials availability} 

The data that support the findings of this study are available from the corresponding author upon request.
 
\bibliographystyle{apalike}
\bibliography{paper}

\end{document}